\documentclass[prl,superscriptaddress,twocolumn,footinbib]{revtex4-2}

 \usepackage{empheq}
\usepackage[colorlinks=true,urlcolor=blue]{hyperref}
\usepackage{graphicx}%
\usepackage{multirow}%
\usepackage{amsmath,amssymb,amsfonts}%
\usepackage{amsthm}%
\usepackage{mathrsfs}%
\usepackage[title]{appendix}%
\usepackage{xcolor}%
 \usepackage{textcomp}%
\usepackage{booktabs}%
\usepackage{algorithmicx}%
\usepackage{algpseudocode}%
\usepackage{listings}%

\newcommand{\R}{\mathbb{R}}
\newcommand{\eps}{\varepsilon}
\newcommand{\half}{\frac{1}{2}}
\newcommand{\oeps}{\frac{1}{\eps}}

\newcommand{\bs}[1]{\boldsymbol{#1}}

\begin{document}

\title{Effective viscosity of a two dimensional passive suspension in a liquid crystal solvent}

	\author{S. Dang}
	\affiliation{Department of Mathematics, The Pennsylvania State University, USA}
 	
	\author{C. Blanch-Mercader}
     \email{carles.blanch-mercader@curie.fr}
	\affiliation{Laboratoire Physico-Chimie Curie, Institut Curie, Universit\' e PSL, Sorbonne Universit\' e, CNRS UMR168, F-75248 Paris, France}

 \author{L. Berlyand}\email{lvb2@psu.edu}
	\affiliation{Department of Mathematics and Huck Institute for Life Sciences, The Pennsylvania State University, USA}







\begin{abstract}
    Suspension of particles in a fluid solvent are ubiquitous in nature, for example, water mixed with sugar or bacteria self-propelling through mucus. Particles create local flow perturbations that can modify drastically the effective (homogenized) bulk properties of the fluid.  Understanding the link between the  properties of particles and the fluid solvent, and the  effective  properties of the medium is a classical problem in fluid mechanics. Here we study a special case of a two dimensional model of a suspension of undeformable particles in a liquid crystal solvent. In the dilute regime, we calculate asymptotic solutions of the perturbations of the velocity and director fields and derive an explicit formula for an effective shear viscosity of the liquid crystal medium. Such effective shear viscosity increases linearly with the area fraction of particles, similar to Einstein formula but with a different prefactor. We provide explicit asymptotic formulas for the dependence of this prefactor on the material parameters of the solvent. Finally, we identify a case of decreasing the effective viscosity by increasing the magnitude of the shear-flow alignment coefficient of the liquid crystal solvent.  
\end{abstract}

\maketitle





\section{Introduction:}\label{Sec:1}

Liquid crystals are anisotropic fluids where particles align on average along a common direction, called the director field \cite{de1993physics}. In nematic phases, the director field is equivalent to its opposite and therefore characterises an axis in space. Due to their unique properties, liquid crystals are of considerable technological importance and have attracted interest in a broad range of fields. Some recent examples are from ferroelectric liquid crystals to thin liquid-crystal films or from active nematics to biological physics \cite{lopez2011drops,marchetti2013hydrodynamics,saw2018biological,blanc2023helfrich}. 

The motion of colloidal particles can be influenced by properties of the liquid crystal solvent \cite{stark2001physics}. For example, the director field in the surrounding region of particles can be distorted due to the anchoring of the director field on the interface of particles. These distortions can, for instance, induce additional long-ranged interactions between particles \cite{brochard1970theory,ramaswamy1996power,poulin1997novel,ruhwandl1997long,lubensky1998topological}. Moreover, the motion of colloidal particles can be used to measure rheological properties of a solvent. A method that has been widely used in biological systems like actin or microtubule suspension, \cite{hou1990tracer,dichtl1999colloidal,mason2000rheology,lin2007viscoelastic,zhang2018interplay,velez2024probing}, bacteria suspensions \cite{valeriani2011colloids,peng2016diffusion,lagarde2020colloidal,figueroa2022non} or tissues \cite{campas2014quantifying,guillamat2022integer,barbazan2023cancer,vian2023situ}. Here we are interested in understanding how the bulk properties of liquid crystals are modified by a suspension of particles. 

Several mathematical approaches have been developed for coarse-graining microscopic models. Among them, homogenization theory offers mathematical tools to deal with well-separated scales both numerically and analytically.  In the context of fluid mechanics, this line of research dates back to the seminal work by Einstein 1905~\cite{Ein1906} who calculated the homogenized viscosity of a dilute suspension of particles in Stokesian fluids. Subsequently, this work was extended to non-dilute regimes, requiring the inclusion of pairwise particle interactions~\cite{BatGre1972}. Since then, numerous extensions of the problem have been explored, including suspensions of particles with different shapes \cite{duerinckx2021corrector,duerinckx2022quantitative,duerinckx2024semi} or suspensions with particles that self-propel \cite{haines2008effective,haines2009three,haines2012effective,girodroux2023derivation}. Other approaches such as the BBGKY hierarchy has been widely used to coarse-grain assemblies of self-propelled particles \cite{bertin2006boltzmann,kruse2006dynamics,baskaran2008enhanced,baskaran2010nonequilibrium,doubrovinski2010self,chou2015active,bertin2015comparison}. 
 
Various experimental and theoretical techniques have been developed to measure the effective viscosities of liquid crystals. The pioneering work of Miesowicz \cite{Mie1946} first reported on the anisotropy of viscosity coefficients in pure liquid crystals under shear flow. In \cite{KneSchSch1991,HeuKneSch1992}, the authors computed the frictional drag on a sphere moving through a liquid crystal solvent with a uniform director alignment parallel to the flow direction at infinity. More recently, \cite{Oza2010} introduced the shear horizontal wave method, which measures an effective viscosity averaged over the depth of the liquid crystal solvent, offering a novel experimental approach to obtain the bulk viscosity.

In this work, we first introduce a 2D mathematical model for describing a suspension of particles immersed in a liquid crystalline medium based on the Ericksen-Leslie model. Next, we perform a two-scale expansion to obtain the local problem for a single particle, which is known in homogenization theory as the cell problem.  Then, we determine the effective viscosity in the dilute regime up to first order in area fraction. Finally, we discuss a case of decreasing the effective viscosity by changing parameters of the liquid crystal solvent. 

\section{Ericksen-Leslie Model for Suspensions}\label{Sec:2}

\begin{figure}[h!]
        \centering
        \includegraphics[width=\columnwidth]{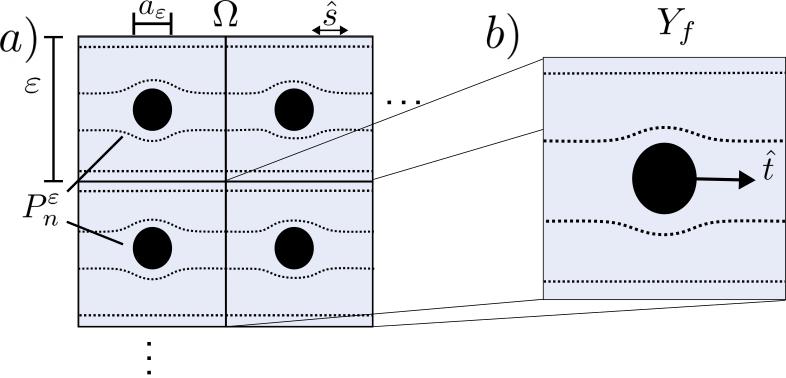}
        \caption{Schematic of a periodic suspension of particles in a liquid crystal. a) $\eps$-periodic square array of undeformable identical discs (black) $\mathcal{P}^\eps=\bigcup_n \mathcal{P}_n^\eps$. The radii of the particles is $a^{\eps}$. The particles are immersed in a liquid crystalline medium occupying a domain $\Omega^\eps = \Omega \setminus\mathcal{P}^\eps$ with $\Omega\subset \R^2$ being the domain of the container. The vector $\hat{s}=\{\cos{(\theta_s)},\sin{(\theta_s)}\}$ is the anchoring on the boundary of the container $\partial\Omega$ with phase $\theta_s$. b) Schematic of the unit cell problem on domain $Y_f$.
        The dimensions of the periodic cell are rescaled such that the size of the domain $Y_f$ is $1$ and the size of the particle is $a$. The vector $\hat{t}=\{\cos{(\theta_t)},\sin{(\theta_t)}\}$ is the anchoring of the director on the particle boundary $\partial\mathcal{P}^\eps$ with phase $\theta_t$. Dotted black lines represent the director field $\bs n^\eps$.  }
        \label{fig:fig1} 
    \end{figure}

In this section, we present the theoretical framework for describing a suspension of solid particles in a liquid crystal. This suspension is modeled by an $\eps$-periodic square array of undeformable identical discs $\mathcal{P}^\eps=\bigcup_n \mathcal{P}_n^\eps$ immersed in a liquid crystaline medium occupying a domain $\Omega^\eps = \Omega \setminus\mathcal{P}^\eps$ with $\Omega\subset \R^2$, see Fig.~\ref{fig:fig1}a. The parameter $\eps$ is the spacing between centers of particles (i.e. period of the square array) and the parameter $a^\eps$ is the radius of the discs.  In the subsequent sections, 
we consider the dilute regime by taking the limit $a^\eps/\eps\to 0$.  Note that because the domain $\Omega$ is divided into an $\eps$-periodic array, it can be partitioned into a periodic array of cells where each cell $Y_f$ contains a single particle $\mathcal{P}^\eps_n$, see Fig.~\ref{fig:fig1}b.

While, in the literature, there are several descriptions for the liquid crystal solvent, in this work, we choose the Ericksen-Leslie model~\eqref{eq: ericksen_leslie_1}-\eqref{eq: ericksen_leslie_2_2} because it is amenable to analytical studies. The liquid crystal is described by a velocity field $\bs u^\eps$ and a director field $\bs n^\eps$. Because the liquid crystal has nematic symmetry, the system of continuum equations governing the hydrodynamics of the liquid crystal are invariant under the transformation $\bs n^\eps\rightarrow -\bs n^\eps$ and it takes the form
%
  \begin{empheq}[left=\empheqlbrace]{align}
        &\nabla \cdot \bs \sigma^\eps = 0 & x \in \Omega^{\eps}\label{eq: ericksen_leslie_1}\\
        &\nabla \cdot \bs u^\eps = 0  & x \in \Omega^{\eps}\label{eq: ericksen_leslie_3}\\
        &\partial_t \bs n^\eps + (\bs u^\eps \cdot \nabla) \bs n^\eps + A(\bs u^\eps) \cdot \bs n^\eps &\nonumber\\
        &=\Gamma \bs h^\eps - \nu D(\bs u^\eps)\cdot \bs n^\eps & x \in \Omega^{\eps} \label{eq: ericksen_leslie_2}\\
         &|\bs n^\eps| = 1 & x \in \Omega^{\eps}\label{eq: ericksen_leslie_2_2}
    \end{empheq}
where $\bs \sigma^\eps$ is the total stress tensor \cite{de1993physics} of the form
\begin{eqnarray}
    \bs \sigma^\eps(x) =&  2\eta D(\bs u^\eps)  - p^\eps +\dfrac{\nu}{2} (\bs n^\eps \bs h^\eps + \bs h^\eps \bs n^\eps) \nonumber \\
    &+ \half (\bs n^\eps \bs h^\eps - \bs h^\eps \bs n^\eps) - K(\nabla \bs n^\eps)(\nabla \bs n^\eps)\label{eq:total_stress}
\end{eqnarray}
In the limiting case where inertial effects are negligible and in the absence of external forces, the conservation of momentum reduces to the force balance equation \eqref{eq: ericksen_leslie_1}. In Eq.~\eqref{eq:total_stress}, the first term on the right hand side (RHS) is the viscous stresses with the shear viscosity $\eta$, where $D(\bs u^\eps)$ is the symmetric part of the velocity gradient tensor $D(\bs u^\eps) = (\partial_\alpha \bs u^{\eps}_\beta + \partial_\beta \bs u^\eps_\alpha)/2$. The second term $p^\eps$ is the pressure field, which acts as a Lagrange multiplier to enforce the incompresibility constraint~\eqref{eq: ericksen_leslie_3}.  The next three terms are stresses due to distortions of the director field which depend on the field $\bs h^\eps=-\delta \mathcal{F}_d/\delta \bs n^\eps$ often called the molecular field in the context of liquid crystals \cite{de1993physics}.  Here $\delta/\delta \bs n^\eps$ refers to the functional derivative of $\mathcal{F}_d$ which is the part of the total free-energy that depends on distortions of the director field.  Here, $\mathcal{F}_d$ is given by the Frank free-energy in the one-constant approximation \cite{de1993physics}
\begin{equation}
    \mathcal{F}_d = \int\limits_{\Omega^\eps} \left(\dfrac{K}{2}(\nabla \bs n^\eps)^2 + \lambda^\eps |\bs n^\eps|^2\right) \, \text{d}a\label{eq:total_freeneergy}
\end{equation}
The elastic coefficient $K$ is the reduced Frank constant in two dimensions and $\text{d}a$ is an infinitesimal element of area. The field $\lambda^\eps$ is a Lagrange multiplier that enforces the constraint~\eqref{eq: ericksen_leslie_2_2}, which means that the liquid crystal is deep into a nematic phase.  Using Eq.~\eqref{eq:total_freeneergy}, the expression of the molecular field becomes
\begin{equation}
    \bs h^\eps := -\dfrac{\delta \mathcal{F}}{\delta \bs n^\eps} = K\Delta \bs n^\eps -\lambda^\eps \bs n^\eps,\label{eq:molecular_field_def}
\end{equation}

The dynamics of the director field are governed by Eq.~\eqref{eq: ericksen_leslie_2}.  The terms on the left hand side (LHS) of Eq.~\eqref{eq: ericksen_leslie_2} represent the co-rotational convective time derivative of $\bs n^\eps$, where $A(\bs u^\eps)$ is the anti-symmetric part of the velocity gradient tensor $A(\bs u^\eps) = (\partial_\alpha \bs u^{\eps}_\beta - \partial_\beta \bs u^\eps_\alpha)/2$. The first term on the RHS of Eq.~\eqref{eq: ericksen_leslie_2} is the elastic torque associated with perturbation of the director field, where $\Gamma$ is the inverse of the rotational viscosity.  The second term on the RHS in Eq.~\eqref{eq: ericksen_leslie_2} couples the dynamics of $\bs n^{\eps}$ to the local shear flows with the parameter $\nu$ called the shear-flow alignment coefficient. This coefficient controls two regimes of the dynamics of the director field: The tumbling regime $|\nu|<1$, where the director field rotates when subjected to shear flows, and the aligning regime $|\nu|>1$, where the director field aligns when subjected to shear flows. It follows from the Onsager relations that the coefficient $\nu$ in Eq.~\eqref{eq:total_stress} and in Eq.~\eqref{eq: ericksen_leslie_2} are equal \cite{de1993physics}. 

As shown in Appendix A, Eq.~\eqref{eq: ericksen_leslie_2} can be further simplified by representing the director field as \begin{equation}
\bs n^\eps=\{\cos(\psi^\eps),\sin(\psi^\eps)\}\label{eq:definition_psi}  
\end{equation}
where $\psi^\eps$ is the angle of the director field with respect to an arbitrary axis. Note that this expression automatically satisfies the constraint~\eqref{eq: ericksen_leslie_2_2}.  Then, the continuum equations ~\eqref{eq: ericksen_leslie_1}-\eqref{eq: ericksen_leslie_2_2} become:
\begin{empheq}[]{align}
        &\nabla \cdot \bs \sigma^\eps = 0 & x \in \Omega^\eps\label{eq: ericksen_leslie_psi_1}\\
        &\nabla \cdot \bs u^\eps = 0& x \in \Omega^\eps\label{eq: ericksen_leslie_psi_4}\\
        &\partial_t  \psi^\eps + (\bs u^\eps \cdot \nabla)  \psi^\eps +\bs n^\eps_\perp A(\bs u^\eps)\bs n^\eps & \nonumber\\&\hspace{10pt}= \Gamma  h^\eps_\perp-\nu \bs n^\eps_\perp D(\bs u^\eps) \bs n^\eps   & x \in \Omega^\eps\label{eq: ericksen_leslie_psi_2}\\
        &  \Gamma h^\eps_\parallel = \nu \bs n^\eps D(\bs u^\eps) \bs n^\eps& x \in \Omega^\eps\label{eq: ericksen_leslie_psi_3}
\end{empheq}
where 
\begin{align}
    &h_\perp^\eps = \bs n_\perp^\eps \cdot \bs h^\eps=K\Delta \psi ^\eps  \label{eq:he_perp}\\
    &h_\parallel^\eps = \bs n^\eps \cdot \bs h^\eps= -K|\nabla  \psi^\eps|^2 - \lambda^\eps\label{eq:he_parallel}
\end{align}
are the perpendicular and parallel components of $\bs h^\eps$ with respect to $\bs n^\eps_\perp=\{-\sin(\psi^\eps),\cos(\psi^\eps)\}$ and $\bs n^\eps$ respectively. Since $\lambda^\eps$ is a Lagrange multiplier, then $h^\eps_\parallel$ in Eq.~\eqref{eq:he_parallel} becomes a Lagrange multiplier that is fixed through Eq.~\eqref{eq: ericksen_leslie_psi_3}.

Next, we introduce boundary conditions for the system \eqref{eq: ericksen_leslie_psi_1}-\eqref{eq: ericksen_leslie_psi_3}. Here, there are two boundaries: the boundary of the particles $\partial\mathcal{P}^\eps=\bigcup_n \partial \mathcal{P}_n^\eps$ and the exterior boundary of the container $\partial \Omega$.  The set of boundary conditions reads 
\begin{empheq}[]{align}
        &D(\bs u^\eps) = 0 &  x \in \mathcal{P}^\eps\label{eq:BC_1}\\
        & \bs u^\eps = \bs E \cdot \bs x &  x \in \partial \Omega~\label{eq:BC_2}\\
        &K(\hat{\bs N} \cdot \nabla) \psi^\eps = W\sin\left( 2 (\theta_t(\bs x) - \psi^\eps)\right) &  x \in  \partial \mathcal{P}^\eps \label{eq:BC_5}\\
        &\psi^\eps=\theta_s(\bs x) &  x \in \partial \Omega~\label{eq:BC_6}
\end{empheq}
The rigidity condition~\eqref{eq:BC_1} extends the velocity $\bs u^{\eps}$ to a constant inside $\mathcal{P^\eps}$ and so that at the particle boundary $\partial \mathcal{P}^\eps$ the velocity field is continuous.  
At the container boundary, we impose in Eq.~\eqref{eq:BC_2} that the velocity is a linear shear flow given by $ \bs E \cdot \bs x$, where $\bs E$ is a symmetric traceless matrix. In addition, the director field on the particle boundary satisfies the torque balance equation~\eqref{eq:BC_5}, where the term on the RHS of ~\eqref{eq:BC_5} corresponds to the torques generate by misalignment of the director field with respect to a preferred anchoring $\hat{\bs t}=\{\cos(\theta_t(\bs x)),\sin(\theta_t(\bs x))\}$ with phase $\theta_t(\bs x)$, Fig.~\ref{fig:fig1}b. The strength of the anchoring alignment is controlled by the parameter $W$ in~\eqref{eq:BC_5}. Vector $\hat{\bs N}$ is the outward normal to the particle boundary. Likewise, on the container boundary in~\eqref{eq:BC_6}, we impose that the director field aligns with a preferred anchoring $\hat{\bs s}=\{\cos(\theta_s(\bs x)),\sin(\theta_s(\bs x))\}$ with a fixed given phase $\theta_s(\bs x)$, Fig.~\ref{fig:fig1}a.

Note that the model \eqref{eq: ericksen_leslie_psi_1}-\eqref{eq:BC_6} depends on the following set of parameters: the size of a cell $\eps$, the radii of discs $a^\eps$, and the size of the container $|\Omega|$ which does not depend on $\eps$, the shear viscosity $\eta$ and the rotational viscosity $1/\Gamma$, the elastic coefficients $K$ and $W$, two given phases $\theta_s(\bs x)$ on the boundary of the container and $\theta_t(\bs x)$ on the boundary of the particles, the shear flow rate $\bs E$, which has two independent components, and the dimensionless shear-flow alignment coefficient $\nu$. In the following, we consider the special case where the phases are equal and constant $\theta_s=\theta_t=\text{\emph{constant}}$ and without loss of generality this constant phase is set to zero.  Moreover, we choose an extensional shear flow at infinity for shear strength $\gamma > 0$ 
\begin{equation}
    \bs E = \gamma\begin{pmatrix}
        1 & 0\\
        0 & -1
    \end{pmatrix}\label{eq:extensional_flow}
\end{equation}
Our focus is on steady state solutions and therefore we consider time-independent physical fields and disregard the time derivative in~\eqref{eq: ericksen_leslie_psi_2}. 

Note that in the absence of particles, the liquid crystal in the configuration $\psi^\eps=0$ is linearly stable to long-wavelength perturbations of the phase when $\nu \gamma <0$, \cite{de1993physics}.  In this work, we choose to consider the range of $\gamma > 0$ and therefore the stable range is $\nu < 0$.


\section{Two-Scale Expansion}\label{Sec:3}

In this section, we perform a two-scale expansion of the solutions $(\bs u^\eps, \psi^\eps)$ to the Eqs.~\eqref{eq: ericksen_leslie_psi_1}-\eqref{eq:BC_6} (see also~\eqref{eq:definition_psi}). We consider the limit $\eps \ll 1$ and derive the so-called \emph{cell problem} which describes the local flows around a particle. We solve the cell problems under two conditions on the range of the parameters $\nu$ and $a^\eps$.  \begin{enumerate}
    \item[{\bf I.}] Two limiting regimes of the shear-flow alignment coefficient, where cell problem can be solved asymptotically: $|\nu| \ll 1$ and $|\nu| \gg 1$. 
    \item[{\bf II.}] The dilute regime where $a^\eps/\eps \to 0$ and interparticle interactions can be neglected.
\end{enumerate}
Two-scale expansion is an asymptotic technique that is useful in constructing approximate solutions for problems with two well-separated scales.  This technique has been used in a wide variety of contexts e.g., \cite{BerRyb2018, BenLioPap2011, MarKhr2008}.  For example, it has been used to study suspensions in the Stokes fluids \cite{All1991} as well as nonhomogeneous materials, including elastic materials and conductive materials \cite{OleShaYos2009, San1980}. In short, we consider the physical fields, $\bs u^\eps$ and $\bs n^\eps$ (equiv. $\psi^\eps$), as vectorial functions of slow varying spatial coordinate $\bs x$ and ``fast'' variable $\bs y=\bs x/\eps$.  This expansion 
 describes two well-separated length scales in the system and the key mathematical idea is to treat $\bs x$ and $\bs y$ as independent variables. The slow variable $\bs x$ represents variations of the physical fields on the scale of the domain $\Omega$ which we normalize to be of order $O(1)$ as $\eps \to 0$.  On the other hand, the fast variable $\bs y$ represents variations of the physical fields on the scale of the interparticle distance of order $O(\eps)$.  Then, the gradient operator writes as:
\begin{equation}
    \nabla \bs f (\bs x,\bs y) = \left(\nabla^x + \oeps \nabla^y \right)\bs f (\bs x,\bs y)\label{eq:example}
\end{equation}
where the superscripts in $\nabla$ denote derivatives with respect to either the slow variable $\bs x$ or the fast variable $\bs y$. We expand the unknown physical fields in a power series in $\eps$ for $\bs u^\eps$, $\psi^\eps$, $p^\eps$, $h_\perp^\eps$ and $h_\parallel^\eps$:

\begin{equation}
    \begin{cases}
        \bs u^\eps(\bs x,\bs y) = \sum_{n=0}^{\infty}\bs u^{(n)}(\bs x,\bs y)\eps^n\\
         \psi^\eps(\bs x,\bs y) = \sum_{n=0}^{\infty}\psi^{(n)}(\bs x,\bs y)\eps^n\\
        p^\eps(\bs x,\bs y) = \sum_{n=-1}^{\infty}p^{(n)}(\bs x,\bs y)\eps^n\\
        h_\perp^\eps(\bs x,\bs y) =\sum_{n=-2}^{\infty} h_\perp^{(n)}(\bs x,\bs y)\eps^n\\
        h_\parallel^\eps(\bs x,\bs y) = \sum_{n=-2}^{\infty}h_\parallel^{(n)}(\bs x,\bs y)\eps^n
    \end{cases}\label{eq:two_scale_expansion}
\end{equation}
The superscripts in the coefficient fields denote the order of the term in the two-scale expansion~\eqref{eq:two_scale_expansion}. Because the domain $\Omega$ is partitioned into an $\eps$-periodic array of cells $Y_f^\eps$, we consider all terms in the RHS of Eq.~\eqref{eq:two_scale_expansion} to have the same $\eps$-periodicity in $\bs y$. 


Substituting Eqs.~\eqref{eq:two_scale_expansion} into the system~\eqref{eq: ericksen_leslie_psi_1}-\eqref{eq: ericksen_leslie_psi_3} and~\eqref{eq:BC_1}-\eqref{eq:BC_6} (see Appendix B to D for calculations), we obtain at the lowest order $\eps^{-2}$ 
\begin{eqnarray}
\psi^{(0)}(\bs x)&=&\theta_s\label{eq:leading_order_terms_1}\\
\bs u^{(0)}(\bs x)&=&\bs E\cdot \bs x\label{eq:leading_order_terms_2}\\
p^{(-1)}&=& h_\perp^{(-2)}= h_\parallel^{(-2)}=0\label{eq:leading_order_terms_3}
\end{eqnarray}
The lowest order terms given by Eqs.~\eqref{eq:leading_order_terms_1}-\eqref{eq:leading_order_terms_3} do not depend on the fast variable $\bs y$ and therefore represent the macroscopic (or coarse-grained) physical fields.

\begin{figure}[h!]
        \centering
        \includegraphics[width=\columnwidth]{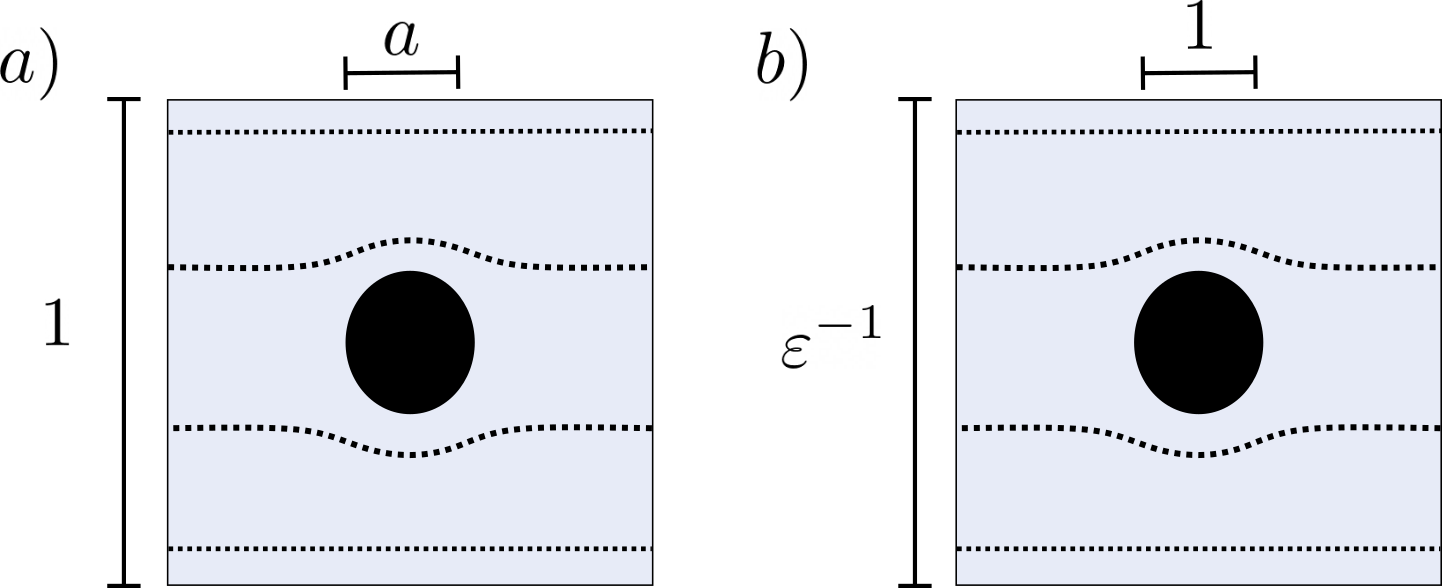}
        \caption{Schematic of the unit cell domain $Y_f$ and the rescaled domain in the dilute limit. a) In the unit cell domain $Y_f$, the particle is of size $a$ in a container of size $1$. In this work, we choose the scaling of $a$ with $\eps$ in Eq.~\eqref{eq:diluteness_parameter}. b) Rescaled domain for the choice of scaling of $a$ with $\eps$ in Eq.~\eqref{eq:diluteness_parameter}. In this domain, the particle is of size $1$ in a container of size $\eps^{-1}$. The dotted curves represents the director field and the black circles represent the particles. }
        \label{fig:fig2} 
    \end{figure}

Due to the presence of particles, local perturbations can be generated on the physical fields, which are accounted for by the next order terms in the two-scale expansions Eqs.~\eqref{eq:two_scale_expansion}. To order $\eps^{-1}$, the perturbations of the director phase and of the two components of the molecular field vanish, $\psi^{(1)}=0$ and $h_\perp^{(-1)}= h_\parallel^{(-1)}=0$ (see Appendix B). Moreover, the perturbations of the velocity and pressure field solve the following ''cell problem'' (see Appendix C and D)
\begin{empheq}[]{align}
        &\eta_1 \partial_{11}^y u_1^{(1)}+\eta_2 \partial_{22} ^y u_1^{(1)}-\partial_1^y\bar{p}^{(0)}=0 &  y \in Y_f\label{eq:PDEu1_1} \\
        & \eta_1 \partial_{11}^y u_2^{(1)}+\eta_2 \partial_{22} ^y u_2^{(1)}-\partial_2^y\bar{p}^{(0)}=0 &  y \in Y_f \label{eq:PDEu1_2} \\
        & \partial_1^y u_1^{(1)}+\partial_2^y u_2^{(1)}=0  &  y \in Y_f \label{eq:PDEu1_3} \\
        & \bs u^{(1)}~\text{periodic} & y\in \partial Y_f \label{eq:PDEu1_4}\\
        & \bs u^{(1)}=-\bs E\cdot \bs y & y\in \partial \mathcal{P} \label{eq:PDEu1_5}
\end{empheq}
where $\eta_1=\eta+(\nu-1)^2/4\Gamma$ and $\eta_2=\eta+(\nu+ 1)^2/4\Gamma$ are two effective viscosities. The subscripts in the derivatives denote the Cartesian coordinates in two dimensions. The pressure field $p^{(0)}$ is expressed in terms of $\bar{p}^{(0)}$, which satisfies the Laplace equation (see Appendix D). Because the ansatz \eqref{eq:two_scale_expansion} is $\eps$-periodic in the fast variable $\bs y$, it is sufficient to find the solution in a single cell $Y_f$ with periodic boundary conditions on $\partial Y_f$~\eqref{eq:PDEu1_4} and the boundary conditions~\eqref{eq:PDEu1_5} on $\partial \mathcal{P}$.  This is the so-called cell problem which depends on the radius of the particles and the material parameters of the liquid crystal solvent, such as the shear-flow alignment coefficient $\nu$.  
We note that since the cell problem only depends on $\bs y$ the domain $Y_f$ is of size $1$ and the particle radius becomes $a$, see Fig.~\ref{fig:fig2}a

The cell problem~\eqref{eq:PDEu1_1}-\eqref{eq:PDEu1_3} describe an anisotropic  fluid with two viscosities $\eta_1$ and $\eta_2$. Because the fluid is incompressible, the velocity field can be expressed in terms of a streamfunction $\Psi^{(1)}$, such that $u^{(1)}_1=-\partial^y_2\Psi^{(1)}$ and $u^{(1)}_2=\partial^y_1\Psi^{(1)}$. This streamfunction is a solution of a generalised (anisotropic) biharmonic equation (see Appendix D) with the general form
\begin{equation}
    \Psi^{(1)}(\bs y)=Re\left[f(y_1+i y_2)+g\left(\frac{y_1}{\sqrt{\eta_1}}+i \frac{y_2}{\sqrt{\eta_2}}\right)\right]\label{eq:stream1}
\end{equation}
where $f$ and $g$ are analytic functions and $y_1$ and $y_2$ are the Cartesian coordinates of the fast variable $\bs y$. 
  By incompressibility, the pressure field $\bar{p}^{(0)}$ satisfies the $\Delta \bar p^{(0)} = 0$. 
In order to solve the cell problem~\eqref{eq:PDEu1_1}-\eqref{eq:PDEu1_3} analytically, we focus on the limiting regime {\bf I} introduced at the beginning of this section  controlled by the parameter $\nu$ (see Appendix E for calculations).

In the limit of small shear-flow alignment $|\nu|\ll 1$, we solve the cell problem by expanding the perturbations of the streamfunction and pressure field as a power series in $\nu$.
\begin{equation}
\begin{cases}
    \Psi^{(1)}(\bs y) &= \sum_{k=0} \nu^k \Psi^{(1,k)}_s(\bs y)\\
    \bar{p}^{(0)}(\bs y) &= \sum_{k=0} \nu^k \bar{p}^{(0,k)}_s(\bs y)
    \label{eq:smallnu_power_series_expansion}
\end{cases}
\end{equation}
This allows us to transform~\eqref{eq:PDEu1_1}-\eqref{eq:PDEu1_3} into a set of PDEs for the functions $\Psi^{(1,k)}(\bs y)$ and $\bar{p}^{(0,k)}( \bs y)$ which describe an isotropic fluid solvent with shear viscosity $\eta+1/4\Gamma$ and forcing terms determined by the previous order in $\nu$ (see Appendix E).  

Next, we consider the dilute limit of the cell problem where the particle radius $a\to 0$.  Since the PDE for $\bs u^{(1)}$ reduces to an isotropic Stokes equation in the limit of $|\nu| \ll 1$, we choose the scaling \cite{All1991}
\begin{equation}
    a^\eps \propto \eps^2, \quad a \propto \eps\label{eq:diluteness_parameter}
\end{equation}  
Later we will show that this choice of scaling for the radius $a$ leads to corrections to the rate of total dissipation for the suspension.

The cell problem on domain $Y_f$ can be rescaled so the radius of the particle is of size $1$ and the domain is of size $\eps^{-1}$, Fig.~\ref{fig:fig2}b.  In the dilute limit ($a\to 0$) this rescaled domain approaches $\R^2 \setminus P$ and we replace the periodic boundary conditions on $Y_f$ \eqref{eq:PDEu1_4} with decay condition at infinity $\bs u^{(1)}\rightarrow 0$ at $\infty$. Then, we solve for $\Psi^{(1)}$ and $\bar{p}^{(0)}$ on the infinite domain. The solution in this infinite domain is an approximation to leading order of the cell problem on a domain of size $1$ with a particle of size $a$, Fig.~\ref{fig:fig2}a. 
Up to order $k=2$ in $\nu$, the particular solution in the unit cell takes the form (see Appendix E)
    \begin{equation}
    \hspace{-5pt}
        \begin{cases}
            \Psi^{(1,0)}_s&=\gamma a^2\left(1-\frac{ a^2}{2r^2}\right)\sin(2\theta)\\
            \Psi^{(1,1)}_s&=\frac{\gamma a^2}{2(1 + 4 \Gamma \eta)}\left(1-\frac{ a^2}{r^2}\right)^2\sin(4\theta)\\
            \Psi^{(1,2)}_s&=\frac{\gamma a^2}{6(1 + 4 \Gamma \eta)^2}\left(2-\frac{5 a^2}{r^2}\right)\left(1-\frac{ a^2}{r^2}\right)^2\sin(6\theta)\\
            \bar{p}^{(0,0)}_s&=-\frac{\gamma a^2(1 + 4 \Gamma \eta)}{\Gamma r^2}\cos(2\theta)\\
            \bar{p}^{(0,1)}_s&=0\\
            \bar{p}^{(0,2)}_s&=-\frac{4\gamma a^2 \eta}{(1 + 4 \Gamma \eta) r^2}\cos(2\theta)
         
        \end{cases}\label{eq:streamfunctionsmallnu}
    \end{equation}
where the Cartesian coordinates $y_1=r\cos{\theta}$ and $y_2=r\sin{\theta}$ are expressed in the polar coordinates $r$ and $\theta$. 

In the limit of large shear-flow alignment $|\nu|\gg 1$, we solve the cell problem in a similar way and expand the perturbations of the streamfunction and pressure field as a power series in $1/\nu$.
\begin{equation}
\begin{cases}
    \Psi^{(1)}(\bs y) &= \sum_{k=0} \nu^{-k} \Psi^{(1,k)}_l(\bs y)\\
    \bar{p}^{(0)}(\bs y) &= \nu^2 \sum_{k=0} \nu^{-k} \bar{p}^{(0,k)}_l(\bs y)
\label{eq:largenu_power_series_expansion}
\end{cases}
\end{equation}
By a similar procedure, up to order $k=2$ in $1/\nu$, the particular solution in the unit cell takes the form (see Appendix E)
    \begin{equation}
        \begin{cases}
            \Psi^{(1,0)}_l&=\gamma a^2\left(1-\frac{ a^2}{2r^2}\right)\sin(2\theta)\\
            \Psi^{(1,1)}_l&=\frac{\gamma a^2}{2}\left(1-\frac{ a^2}{r^2}\right)^2\sin(4\theta)\\
            \Psi^{(1,2)}_l&=\frac{\gamma a^2}{6}\left(2-\frac{5 a^2}{r^2}\right)\left(1-\frac{ a^2}{r^2}\right)^2\sin(6\theta)\\
            \bar{p}^{(0,0)}_l&=-\frac{\gamma a^2}{\Gamma r^2}\cos(2\theta)\\
            \bar{p}^{(0,1)}_l&=0\\
            \bar{p}^{(0,2)}_l&=-\frac{4\gamma a^2 \eta}{ r^2}\cos(2\theta)
        \end{cases}\label{eq:streamfunctionlargenu}
    \end{equation}
where the Cartesian coordinates $y_1=r\cos{\theta}$ and $y_2=r\sin{\theta}$ are expressed in the polar coordinates $r$ and $\theta$. 

In both limiting regimes of $\nu$, the net force and net torque on the particle boundary associated with either perturbation \eqref{eq:streamfunctionsmallnu} or \eqref{eq:streamfunctionlargenu} vanish, and therefore the particle's linear and angular velocity vanish (see Appendix F for calculations).

\section{Dissipation rate and Effective Viscosity}\label{Sec:4}

In this section, we will determine an effective viscosity of a suspension of colloidal particles in a liquid crystal solvent. Using the results from the two-scale expansion in the previous section, we will determine the rate of total dissipation for two systems: a dilute suspension of particles in a liquid crystal solvent and an homogenized liquid crystal. The comparison between the two rates of total dissipation allows us to identify an effective viscosity for the suspension and its dependence on the particle density and material parameters of the liquid crystal solvent.

Finding an equivalent homogenized medium without particles that produces the same rate of total dissipation in a macroscopic volume as a suspension in a solvent is a classical problem in fluid dynamics, \cite{Ein1906, BatGre1972}. In the case of a suspension of rigid particles in a liquid crystal solvent, the rate of total dissipation is equal to the time variation of the total free-energy which takes the form
\begin{equation} 
    \frac{d \mathcal{F}}{dt}=-\int_{\Omega^\eps}2 \eta D(\bs u^\eps):D(\bs u^\eps)+\Gamma \bs h^\eps\cdot \bs h^\eps da\label{eq:Energydissip}
\end{equation}
where  $\bs u^\eps$ is defined in~\eqref{eq:two_scale_expansion}, $\bs h^\eps$ is defined in~\eqref{eq:molecular_field_def}, and the operator $:$ denotes a contraction of the two indices of the symmetric velocity gradient tensor $D(\bs u^\eps)$. In this expression, the first term is the rate of dissipation due to viscous stresses and the second term is the rate of dissipation due to dynamics of the director field, which has no counterpart in Newtonian fluids. 

We first focus on the dissipation due to viscous stresses in Eq.~\eqref{eq:Energydissip}.  We express the area of the container $|\Omega|$ minus the area of the particles of radius $a^\eps$
\begin{equation}
    |\Omega^\eps| = |\Omega| - N_p \pi (a^\eps)^2\label{eq:omega_expansion_1}
\end{equation}
where $N_p = C \eps^{-2}$ is the number of particles, see Fig.~\ref{fig:fig1}a and $a^\eps$ is given by Eq.~\eqref{eq:diluteness_parameter}.  Substituting the two-scale expansion~\eqref{eq:two_scale_expansion} into the first term in the RHS of Eq.~\eqref{eq:Energydissip} to obtain
\begin{align}
\int_{\Omega^\eps}2 \eta D(\bs u^\eps):&D(\bs u^\eps) d\bs x=4\eta\gamma^2 (|\Omega| - N_p \pi (a^\eps)^2) \nonumber\\
&\hspace{-20pt}+\int_{\Omega^\eps}2 \eta D^y(\bs u^{(1)}):D^y(\bs u^{(1)}) d\bs x +{\cal O}(\eps^3)\label{eq:dissipation_expansion_1}
\end{align}
where we used that $\int_{\Omega^\eps}D^y(\bs u^{(1)})da=0$ and $\gamma$ is the shear strength. The first contribution in the RHS in Eq.~\eqref{eq:dissipation_expansion_1} is the rate of dissipation due to the macroscopic shear flow in the solvent. The second term in the RHS in Eq.~\eqref{eq:dissipation_expansion_1} is the rate of dissipation due to the flow perturbations in the surrounding region of particles.  As written, this term has contributions of order $\eps^{n}$ for $n\geq 2$ which comes from the dependence of $\Psi_s^{(1,k)}$ or $ \Psi_l^{(1,k)}$ on the radius of the particle $a$ in Eqs.~\eqref{eq:streamfunctionsmallnu} and~\eqref{eq:streamfunctionlargenu} and the dependence of the domain $\Omega^\eps$ on $a^\eps$.  In the following, we ignore the additional contributions at order $\eps^{n}$ for $n>2$.

Because Eqs.~\eqref{eq:PDEu1_1}-\eqref{eq:PDEu1_5} are linear in the perturbation flows $\bs u^{(1)}$ and $\bs u^{(1)}$ is $\eps$-periodic in $\bs y$, the second term in the RHS in Eq.~\eqref{eq:dissipation_expansion_1} is expressed as the sum of the rate of dissipation by the flow perturbations of a single particle $\mathcal{P}$ in the unit cell, Fig.~\ref{fig:fig2}a. 
\begin{eqnarray}
    &&\int_{\Omega^\eps}2 \eta D^y(\bs u^{(1)}):D^y(\bs u^{(1)}) d\bs x =\nonumber\\&&\hspace{20pt} N_p \eps^2\int_{Y_f} 2\eta D^y(\bs u^{(1)}):D^y(\bs u^{(1)}) d\bs y
\end{eqnarray}
We focus on the dilute regime $\eps \to 0$, ~\eqref{eq:diluteness_parameter}, where inter-particle interactions can be neglected, and approximate the integral on the RHS by using the solutions of the perturbations in the two limiting cases of $\nu$, \eqref{eq:streamfunctionsmallnu} and \eqref{eq:streamfunctionlargenu}, where the integration domain is replaced by $\R^2/\mathcal{P}$. The results obtained take the form
\begin{eqnarray}
    &&-\int_{\Omega^\eps}2 \eta D(\bs u^\eps):D(\bs u^\eps) da= -4\eta \gamma^2 \pi R^2 *\nonumber\\
    &&\left\{ \begin{array}{ll}
       (1+\phi\frac{2\nu^2}{(1+4\alpha)^2}) +{\cal O}(\phi^2)+ {\cal O}(|\nu|^{3}) & \mbox{if $|\nu| \ll 1$}\\
        1+{\cal O}(\phi^2)+{\cal O}(|\nu|^{-1}) & \mbox{if $|\nu| \gg 1$}\end{array} \right. \label{eq:energydisspLC1Final}
\end{eqnarray}



where $\phi=N_p (a^{\eps})^2/R^2=\eps^2/R^2$ is the area fraction of particles in the container $\Omega$, and $\alpha=\Gamma \eta$ is a dimensionless parameter that arises from the ratio between the two viscosities $\eta$ and $1/\Gamma$.

Recall that the limit in $\nu$ is performed before the limit in $a$.
  In a similar way, the rate of dissipation due to dynamics of the director field in  Eq.~\eqref{eq:Energydissip} takes the form 
in the limit $|\nu|\ll 1$,
    \begin{align}
      -\int_{\Omega^\eps} &\Gamma \bs h^\eps\cdot \bs h^\eps da=-\frac{\gamma^2 \pi R^2}{\Gamma}*\label{eq:energydisspLC2Final}\\
      &\left(\nu^2+2\phi\left(1-\frac{(1+8\alpha)\nu^2}{(1+4\alpha)^2}\right)\right) +{\cal O}(\phi^2)+{\cal O}(|\nu|^{3})   \nonumber
  \end{align}
  and in the limit $|\nu|\gg 1$,
  \begin{align}
      -\int_{\Omega^\eps} &\Gamma \bs h^\eps\cdot \bs h^\eps da=\label{eq:energydisspLC2Final}\\
      &-\frac{\gamma^2 \pi R^2}{\Gamma}\left(\nu^2+2\phi \right)+ {\cal O}(\phi^2)+{\cal O}(|\nu|^{-1}) \nonumber
  \end{align}

Adding these two expressions yields the rate of total dissipation in the suspension. In the regime of small $|\nu|\ll 1$, dissipation due to viscous stresses \eqref{eq:energydisspLC1Final} is modified by parameters that control the dynamics of the director field such as $\nu$ and  $\Gamma$. Conversely, dissipation due to the dynamics of the director field \eqref{eq:energydisspLC2Final} is modified by parameters that control the viscous stresses, such as the shear viscosity $\eta$. This shows that the presence of particles modifies the rate of total dissipation of the suspension by coupling the two sources of dissipation in the liquid crystal solvent.

The presence of a particle in a solvent influences the rate of total dissipation in suspensions in several ways. On the one hand, particles create perturbations to the macroscopic fields in a region surrounding the particles, increasing the density of dissipation in the solvent. On the other hand, particles occupy space in the container which reduces the area of the solvent  which reduces the total dissipation per unit time (see the first term on the RHS of Eq.~\eqref{eq:dissipation_expansion_1}). The balance between these opposing contributions determines the first order correction in $\phi$ of the various sources of  dissipation~\eqref{eq:energydisspLC1Final} and \eqref{eq:energydisspLC2Final}.

Next, we determine the rate of total dissipation of an homogenized liquid crystal with effective material parameters. For clarity we denote with the superscript $H$, the physical fields and material parameters of the homogenized medium. A priori, all material parameters of the homogenized medium can differ from the material parameters of the solvent.  In this work we focus on the computation of the effective viscosity $\eta^H$ via the solution of the cell problem~\eqref{eq:PDEu1_1}-\eqref{eq:PDEu1_5}, but we note that the effective rotational viscosity $1/\Gamma^H$ and shear-flow alignment coefficient $\nu^H$ could be computed in a similar manner.  
For simplicity, we set $\Gamma^H=\Gamma$ and $\nu^H=\nu$. The homogenized medium is subjected to the same extensional shear flow and preferred  anchoring at the boundaries of the container as the suspension \eqref{eq:BC_2} and \eqref{eq:BC_6}. The expression of the physical fields of the homogenized medium are determined in Appendix G.

The time variation of the total free-energy in the homogenized liquid crystal is   
\begin{equation}
    -\frac{d \mathcal{F}^H}{dt}=\int_{\Omega}2 \eta^H D(\bs u^H):D(\bs u^H)+\Gamma \bs h^H\cdot \bs h^H da\label{eq:EnergydissipH}
\end{equation}
Substituting the expression of the physical fields of the homogenized medium (see Appendix G), the rate of total dissipation up to ${\cal O}(\eps^3)$ in \eqref{eq:EnergydissipH} takes the form in the two limits of $\nu$
\begin{align}
   \frac{d \mathcal{F}^H}{dt}&=-C_R*\nonumber\\
   &\left\{ \begin{array}{ll}
        \left(1-2\phi \right)+ {\cal O}(\phi^2)+{\cal O}(|\nu|^{3})  & \mbox{if $|\nu| \ll 1$};\\
         \left(1-2\phi\right)+ {\cal O}(\phi^2)+{\cal O}(|\nu|^{-1}) & \mbox{if $|\nu| \gg 1$}.\end{array} \right.\label{eq:energydisspHFinal} 
\end{align}
where the constant $C_R = \left(4\eta^H+\frac{\nu^2}{\Gamma}\right)\gamma^2 \pi R^2$. 

Balancing the sum of Eqs.~\eqref{eq:energydisspLC1Final} and \eqref{eq:energydisspLC2Final} to Eq.~\eqref{eq:energydisspHFinal} allows us to define an effective viscosity up to ${\cal O}(\eps^3)$ for the homogenized medium in the limit $|\nu| \ll 1$:
\begin{align}
  \frac{\eta^H}{\eta}&=        \label{eq:EV_nu_small}\\
        &1+\phi\left(\frac{1+4 \alpha}{2 \alpha}-\frac{\nu^2 (1+8 \alpha)}{8 \alpha^2 (1+4 \alpha)}\right) +{\cal O}(\phi^2)+{\cal O}(|\nu|^{3})  \nonumber
\end{align}
and in the limit $|\nu| \gg 1$:
\begin{equation}
    \frac{\eta^H}{\eta}=1+2\phi\left(1+\frac{1}{\nu^2}\right)  + {\cal O}(\phi^2)+{\cal O}(|\nu|^{-1})\label{eq:EV_nu_large}
\end{equation}
Note that in the limit $\Gamma\rightarrow\infty$, we recover the well-known viscosity of a two-dimensional suspension in a Stokesian fluid $\eta^H/\eta=1+2\phi+{\cal O}(\phi^2)$. The effective viscosity \eqref{eq:EV_nu_small} and \eqref{eq:EV_nu_large} depends linearly on the area fraction of particle $\phi$ and the dimensionless parameters $\alpha $ and $\nu$. 

For sufficiently small values of $|\nu|\ll 1$, the effective viscosity \eqref{eq:EV_nu_small} depends on $\alpha $ even in the case of vanishing $\nu$. At a fixed area fraction, the effective viscosity can be reduced by modifying properties of the liquid crystal solvent, such as increasing the parameter $|\nu|$ or the dimensionless parameters $1/\alpha$. For sufficiently large values $|\nu|\gg 1$, we recover the Einstein formula for a Stokesian fluid in Eq.~\eqref{eq:EV_nu_large}, $\eta^H/\eta=1+2\phi+{\cal O}(\phi^2)$. 

\begin{figure}[h!]
        \centering
        \includegraphics[width=8.6 cm]{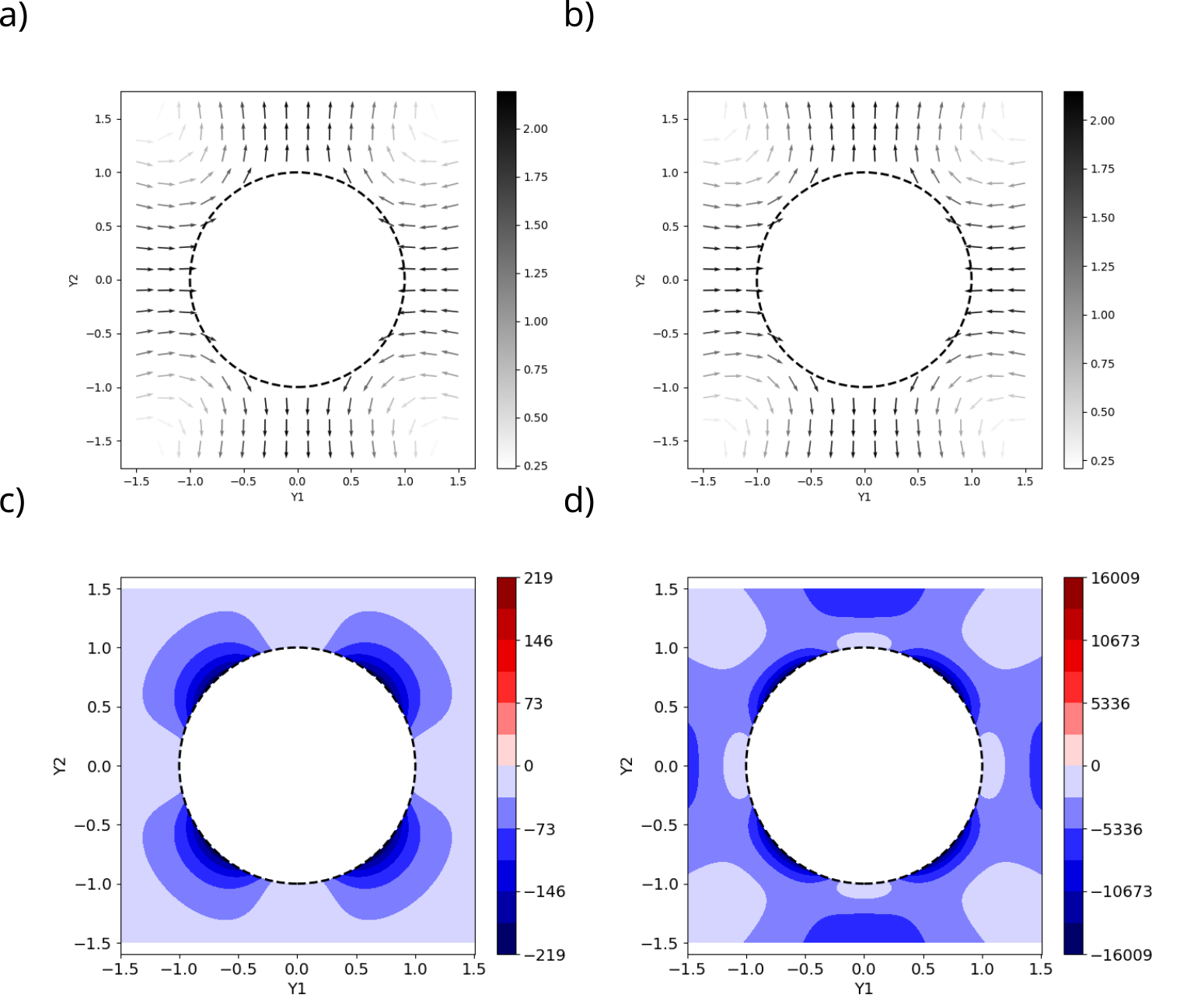}
        \caption{Flow perturbation $\bs u^{(1)}$ in the limit $|\nu| \ll 1$ (tumbling phase) (a) and in the limit $|\nu| \gg 1$ (aligning phase) (b).  For a background shear flow that extends in the horizontal axis and contracts in the vertical axis.  Colormap of the density of energy dissipation per unit time (integrand in~\eqref{eq:Energydissip}) in the limit $|\nu| \ll 1$ (c) and in the limit $|\nu| \gg 1$ (d).  Note that locally, the density of energy dissipation remains negative (shown in blue).  In panels a-d, the parameter set is: $\gamma = 2$, $1/\Gamma = 10$, $\eta = 1$ and $a=1$. In panels a and c, we used $\nu=0.01$ and in panels b and d, we used $\nu=10$. In panels c and d, the dashed circle represents the boundary of a particle with radius $a$.           
        }
        

        \label{fig:fig3} 
    \end{figure}

Previous studies of suspensions of active swimmers in a Stokesian fluid have shown both experimentally and theoretically that the homogenized viscosity (i.e., effective viscosity) of the suspension can be decreased by increasing the area fraction $\phi$ \cite{haines2008effective,haines2009three,haines2012effective,girodroux2023derivation, gachelin2013non,lopez2015turning,clement2016bacterial,martinez2020combined}. This effect is captured by a negative term due to activity in the classical Einstein formula.  Observe that there is a negative term in~\eqref{eq:EV_nu_small} (second order in $\nu$ and first order in $\phi$). This term can cause a decrease of the effective viscosity by either increasing the parameter $|\nu|$ or decreasing the parameter $\alpha$ at a fixed area fraction $\phi$. However, it cannot lead to a decrease of the effective viscosity ($\eta^H < \eta$) by increasing the area fraction $\phi$ in the parameter regimes considered above ($|\nu|\ll 1$ and $|\nu|\gg 1$). To illustrate that in our case $\eta^H >0$, we computed the density of energy dissipation per unit time in the solvent (i.e., the integrand in Eq.~\eqref{eq:Energydissip}), see Fig~\ref{fig:fig3}, and found that it is overall negative in the two regimes of $|\nu|$.  As the area fraction increases the suspension dissipates more energy per unit time, leading to an increase in the effective viscosity ($\eta^H > \eta$), as shown in Eqs.~\eqref{eq:EV_nu_small} and \eqref{eq:EV_nu_large}. An interesting open question is whether there exists an intermediate parameter regime in $0<|\nu|< \infty$ where this effective viscosity is negative.

\section{Discussion}


In this work, we obtained an effective viscosity for a dilute suspension of solid particles in a liquid crystal solvent.  A key aspect of our analysis is the consideration of two limits: $|\nu| \ll 1$ and $|\nu| \gg 1$ (shear-flow alignment) and $a^\eps / \eps \to 0$ (diluteness parameter).  
The order in which these limits are taken can be critical, as it could affect the resulting behavior of the macroscopic flow and the bulk properties of the homogenized liquid crystal.  While $|\nu| \ll 1$ and $|\nu| \gg 1$ are regular perturbations of a material parameter of the liquid crystal solvent, the diluteness parameter $a^\eps / \eps $ is a geometric parameter which affects the solution at both microscopic and macroscopic scales and thus may not commute with the former limit in $\nu$. 
 By performing the limit in $\nu$ first, we obtain an isotropic Stokes Eqs.~\eqref{eq:nu_expansion_order_0}-\eqref{eq:nu_expansion_order_2}  and~\eqref{eq:largenu_expansion_order_0}-\eqref{eq:largenu_expansion_order_2} for which the scaling of $a^\eps / \eps$ for the dilute regime are known~\cite{All1991}.  Understanding the relationship between $\nu$ and $a^\eps/\eps$ deserves further attention in the future. 



We would like to acknowledge that we learned of formula \eqref{eq:stream1} from J. King (University of Nottingham) and we are grateful to him. The work of L. B. was partially supported by NSF grants DMS-2404546 and PHY-2140010.  S. D. was partially supported by NSF grant PHY-2140010.
    
\bibliography{apssamp}
\onecolumngrid
\appendix
\section{Appendix A: The derivation of the molecular field}
In this Appendix, we derive Eqs.~\eqref{eq: ericksen_leslie_psi_2}-\eqref{eq:he_perp}, which respectively express the dynamics of the director field $\bs n^\eps$ and the components of the molecular field $\bs h^\eps$ in terms of the phase $\psi^\eps$ of the director field.

In two dimensions, the director field can be expressed as $\bs n^\eps=\{\cos(\psi^\eps),\sin(\psi^\eps)\}$. Inserting this expression into Eq.~\eqref{eq:molecular_field_def}, one obtains
\begin{equation}
    \bs h^\eps = K\Delta \bs n^\eps-\lambda^\eps \bs n^\eps = \bs n^\eps\left(-K|\nabla \psi^\eps|^2-\lambda^\eps\right)+\bs n^\eps_\perp\left(K\Delta\psi^\eps\right)\label{eq:appA5}
\end{equation}
where $\bs n^\eps_\perp=\{-\sin(\psi^\eps),\cos(\psi^\eps)\}$ is a vector perpendicular to $\bs n^\eps$. Therefore the components of the molecular field are
\begin{align}
    &h_\perp^\eps = \bs n_\perp^\eps \cdot \bs h^\eps=K\Delta \psi ^\eps  \label{eq:appA3}\\
    &h_\parallel^\eps = \bs n^\eps \cdot \bs h^\eps= -K|\nabla  \psi^\eps|^2 - \lambda^\eps\label{eq:appA4}
\end{align}
Note that the Lagrange multiplier $\lambda^\eps$ associated with the constraint  $|\bs n^\eps|^2=1$ can be re-expressed such that $h_\parallel^\eps$ becomes a Lagrange multiplier.

Inserting the above expression of the director field into the Ericksen-Leslie equation \eqref{eq: ericksen_leslie_2}, one obtains
\begin{equation}
    \bs n^\eps_\perp\left(\partial_t  \psi^\eps + \bs u^\eps \cdot \nabla  \psi^\eps +\bs n^\eps_\perp A(\bs u^\eps)\bs n^\eps\right) = \Gamma  \bs h^\eps-\nu D(\bs u^\eps) \bs n^\eps
\end{equation}
and its projections in the direction parallel and perpendicular to the director field takes the form
\begin{align}
    &\partial_t  \psi^\eps + \bs u^\eps \cdot \nabla  \psi^\eps +\bs n^\eps_\perp A(\bs u^\eps)\bs n^\eps = \Gamma  h^\eps_\perp-\nu \bs n^\eps_\perp D(\bs u^\eps) \bs n^\eps\label{eq:appA1}  \\
        &  \Gamma h^\eps_\parallel = \nu \bs n^\eps D(\bs u^\eps) \bs n^\eps\label{eq:appA2}
\end{align}
which correspond to Eqs.~\eqref{eq: ericksen_leslie_psi_2} and \eqref{eq: ericksen_leslie_psi_3} in the main text. The dynamics of the phase $\psi^\eps$ are governed by Eq.~\eqref{eq:appA1}, where $h_\perp^\eps$ is given by Eq.~\eqref{eq:appA3}. The Lagrange multiplier $h_\parallel^\eps$ is fixed by Eq.~\eqref{eq:appA2}.

\section{Appendix B: Two-scale expansion of the equations for director}

In this Appendix, we perform the two-scale expansion for Eqs.~\eqref{eq: ericksen_leslie_psi_2}-\eqref{eq:he_perp} as well as for the boundary conditions~\eqref{eq:BC_5}-\eqref{eq:BC_6}. For more details on the two-scale expansion, we refer the reader to 
the main text and to references \cite{BerRyb2018, BenLioPap2011}.

Inserting the relations  \eqref{eq:two_scale_expansion} into Eqs.~\eqref{eq: ericksen_leslie_psi_2}-\eqref{eq:he_perp} and Eqs.~\eqref{eq:BC_5}-\eqref{eq:BC_6} and treating the derivative as in Eq.~\eqref{eq:example}, one obtains to the lowest order in $\eps$ that
\begin{empheq}[left=\empheqlbrace]{align}
 \eps^{-2}:~~ &\Gamma h_{\perp}^{(-2)}=0 &  x \in  \Omega^\eps\label{eq:he_expansion_21}\\
   \eps^{-2}:~~ & h_{\perp}^{(-2)}=K \Delta^y \psi^{(0)}&  x \in  \Omega^\eps\label{eq:he_expansion_22}\\
   \eps^{-2}:~~ &\Gamma h_{\parallel}^{(-2)}=0  &  x \in  \Omega^\eps\label{eq:he_expansion_23}\\
   \eps^{-1}:~~ & K(\hat {\bs N} \cdot \nabla^y) \psi^{(0)} = 0 &  x \in \partial \mathcal{P}^\eps \label{eq:anchoring_BC_psi_11}\\
    \eps^{0}:~~& \psi^{(0)}=0 &  x \in \partial \Omega
    \label{eq:anchoring_BC_psi_12}
\end{empheq} 
The solution of this set of PDEs is 
\begin{equation}
    \psi^{(0)} = 0
\end{equation}

The corrections to the next order in $\eps$ of Eqs.~\eqref{eq: ericksen_leslie_psi_2}-\eqref{eq:he_perp} and Eqs.~\eqref{eq:BC_5}-\eqref{eq:BC_6} are
\begin{empheq}[left=\empheqlbrace]{align}
 \eps^{-1}:~~ &\Gamma h_{\perp}^{(-1)}=\bs n^{(0)}_\perp A^y(\bs u^{(0)})\bs n^{(0)}+\nu \bs n^{(0)}_\perp D^y(\bs u^{(0)}) \bs n^{(0)} &  x \in  \Omega^\eps\label{eq:he_expansion_11}\\
   \eps^{-1}:~~ & h_{\perp}^{(-1)}=K\Delta^y \psi^{(1)} &  x \in  \Omega^\eps\label{eq:he_expansion_12}\\
   \eps^{-1}:~~ &\Gamma h_{\parallel}^{(-1)}=\nu \bs n^{(0)} D^y(\bs u^{(0)}) \bs n^{(0)}  &  x \in  \Omega^\eps\label{eq:he_expansion_13}\\
   \eps^{0}:~~ & K(\hat {\bs N} \cdot \nabla^y) \psi^{(1)} = 0 &  x \in \partial \mathcal{P}^\eps \label{eq:anchoring_BC_psi_01}\\
    \eps^{1}:~~& \psi^{(1)}=0 &  x \in \partial \Omega
    \label{eq:anchoring_BC_psi_02}
\end{empheq}
The advection in Eq.~\eqref{eq: ericksen_leslie_psi_2} does not appear at order $\eps^{-1}$ because the lowest order term of the director's phase $\psi^{(0)}$ is independent on the fast variable.

In Appendix D, we will show that the lowest order term of the velocity field $u^{(0)}$ is only a function of the slow variable $\bs x$. Note that in this case Eqs. \eqref{eq:he_expansion_11} and \eqref{eq:he_expansion_13} simplify to $h_{\perp}^{(-1)}=h_{\parallel}^{(-1)}=0$, and therefore the solution of the previous set of PDEs is 
\begin{equation}
    \psi^{(1)} = 0
\end{equation}

Finally, the corrections to order $\eps^{0}$ of Eqs.~\eqref{eq: ericksen_leslie_psi_2} and \eqref{eq: ericksen_leslie_psi_3} are 
\begin{align}
    \Gamma h_{\perp}^{(0)} &=\bs n^{(0)}_\perp \left( A^x(\bs u^{(0)})  +  A^y(\bs u^{(1)}) \right)\bs n^{(0)}+\nu \bs n^{(0)}_\perp \left(D^x(\bs u^{(0)})+ D^y(\bs u^{(1)})\right) \bs n^{(0)}\label{eq:he_expansion_01}\\
    \Gamma h_{\parallel}^{(0)}&=\nu \bs n^{(0)} \left(D^x(\bs u^{(0)}) + D^y(\bs u^{(1)})\right) \bs n^{(0)}\label{eq:he_expansion_02}
\end{align}
In our special case, both the lowest order term and the first correction to the director phase vanish $\psi^{(0)}=\psi^{(1)}=0$ and thereby the advection in Eq.~\eqref{eq: ericksen_leslie_psi_2} does not appear at order $\eps^{0}$. 

\section{Appendix C: Two-scale expansion of the stress}

In this Appendix, we perform the two-scale expansion for the total stress tensor \eqref{eq:total_stress}. The results from Appendix B are used to express the molecular field $\bs h_\eps$ in terms of velocity field gradients. The two-scale expansion of the total stress is performed up to first order corrections in $\eps$.

We start from the total stress in index notation and perform the two-scale expansion in component-wise.  From~\eqref{eq:total_stress}
\begin{equation}
    \sigma^\eps_{ij}(x) =  \eta (\partial_i u_j^\eps + \partial_j u_i^\eps)  + \dfrac{\nu}{2} (n^\eps_i h^\eps_j + h^\eps_i n^\eps_j) + \half (n^\eps_i h^\eps_j - h^\eps_i n^\eps_j) - K(\partial_i n^\eps_j)(\partial_i n^\eps_j)- p^\eps \delta_{ij}~\label{eq:total_stress_index}
\end{equation}

We use the notation $\partial_i^\alpha$ where $i=1,2$ is the vector index and $\alpha=x,y$ denotes either the slow or fast variable respectively.  Correspondingly for a function $u_k^{(l)}$, $k=1,2$ denotes the vector index and $l$ denotes the order of $\epsilon$ in the 
two-scale expansion.  In addition, plugging~\eqref{eq:appA5} into the total stress, the second and third terms on the RHS of~\eqref{eq:total_stress_index} are
\begin{align}
    \dfrac{\nu}{2} (n^\eps_i h^\eps_j + h^\eps_i n^\eps_j) &= \dfrac{\nu}{2} \left(2h_\parallel^\eps n^\eps_i n^\eps_j + h_\perp^\eps(n^\eps_i (n_\perp)^\eps_j + (n_\perp)_i^\eps n_j^\eps)\right)\label{eq:appC1}\\
    \half (n^\eps_i h^\eps_j - h^\eps_i n^\eps_j)&=\half  h_\perp^\eps (n^\eps_i (n_\perp)^\eps_j - (n_\perp)_i^\eps n^\eps_j) \label{eq:appC2}
\end{align}
which together with the solution $\psi^{(0)}=0$ from Appendix B, simplifies~\eqref{eq:total_stress_index} at order $\eps^{-1}$ to
\begin{align}
    \sigma_{11}^{(-1)} &= 2\eta \partial_1^y u_1^{(0)}+ \nu h_\parallel^{(-1)} - p^{(-1)}  \label{eq:total_stress_index_he_11_order_-1}\\
    \sigma_{22}^{(-1)} &= 2\eta \partial_2^y u_2^{(0)} - p^{(-1)}  \label{eq:total_stress_index_he_22_order_-1}\\
    \sigma_{12}^{(-1)} &= \eta \left(\partial_1^y u_2^{(0)} + \partial_2^y u_1^{(0)}\right) + \dfrac{\nu+1}{2}h_\perp^{(-1)}\label{eq:total_stress_index_he_12_order_-1}\\
    \sigma_{21}^{(-1)} &= \eta \left(\partial_2^y u_1^{(0)} + \partial_1^y u_2^{(0)}\right) + \dfrac{\nu-1}{2}h_\perp^{(-1)}\label{eq:total_stress_index_he_21_order_-1}
\end{align}
Next, we express Eqs.~\eqref{eq:he_expansion_11} and~\eqref{eq:he_expansion_13} from Appendix B in components and obtain the equations for $h_\parallel^{(-1)}$ and $h_\perp^{(-1)}$ in terms of velocity gradients:
    \begin{align}
        \Gamma h^{(-1)}_\parallel &= \nu \partial_1^y u_1^{(-1)}\label{eq:he_parallel_psi_1}\\
        \Gamma h^{(-1)}_\perp &= \dfrac{(\nu-1)}{2} \partial_1^y u_2^{(0)} + \dfrac{(\nu + 1)}{2} \partial_2^y u_1^{(0)}\label{eq:he_perp_psi_1}
    \end{align}
Finally, substituting~\eqref{eq:he_parallel_psi_1} and~\eqref{eq:he_perp_psi_1} into Eqs.~\eqref{eq:total_stress_index_he_11_order_-1}-\eqref{eq:total_stress_index_he_21_order_-1}, one obtains at order $\eps^{-1}$:
\begin{align}
    \sigma_{11}^{(-1)} &= (2\eta+\dfrac{\nu^2}{\Gamma}) \partial_1^y u_1^{(0)} - p^{(-1)} \label{eq:total_stress_index_he_11_order_-1_v2}\\
    \sigma_{22}^{(-1)} &= 2\eta \partial_2^y u_2^{(0)} - p^{(-1)}\label{eq:total_stress_index_he_22_order_-1_v2}\\
    \sigma_{12}^{(-1)} &= (\eta+\dfrac{\nu^2-1}{4\Gamma}) \partial_1^y u_2^{(0)}+(\eta+\dfrac{(\nu+1)^2}{4\Gamma}) \partial_2^y u_1^{(0)}\label{eq:total_stress_index_he_12_order_-1_v2}\\
    \sigma_{21}^{(-1)} &= (\eta+\dfrac{(\nu-1)^2}{4\Gamma}) \partial_1^y u_2^{(0)}+(\eta+\dfrac{\nu^2-1}{4\Gamma}) \partial_2^y u_1^{(0)}\label{eq:total_stress_index_he_21_order_-1_v2}
\end{align}
Through a similar procedure, one obtains the corrections of the stress components at order $\eps^0$, which take the reduced form: 
\begin{align}
    \sigma_{11}^{(0)} &= (2\eta+\dfrac{\nu^2}{\Gamma}) \left(\partial_1^x u_1^{(0)}+\partial_1^y u_1^{(1)}\right) - p^{(0)} \label{eq:total_stress_index_he_11_order_0_v2}\\
    \sigma_{22}^{(0)} &= 2\eta \left(\partial_2^x u_2^{(0)}+\partial_2^y u_2^{(1)}\right) - p^{(0)}\label{eq:total_stress_index_he_22_order_0_v2}\\
    \sigma_{12}^{(0)} &= (\eta+\dfrac{\nu^2-1}{4\Gamma}) \left(\partial_1^x u_2^{(0)} + \partial_1^x u_2^{(1)}\right)+(\eta+\dfrac{(\nu+1)^2}{4\Gamma}) \left(\partial_2^x u_1^{(0)} + \partial_2^x u_1^{(0)}\right)\label{eq:total_stress_index_he_12_order_0_v2}\\
    \sigma_{21}^{(0)} &= (\eta+\dfrac{(\nu-1)^2}{4\Gamma}) \left(\partial_1^x u_2^{(0)} + \partial_1^x u_2^{(1)}\right)+(\eta+\dfrac{\nu^2-1}{4\Gamma}) \left(\partial_2^x u_1^{(0)} + \partial_2^x u_1^{(0)}\right)\label{eq:total_stress_index_he_21_order_0_v2}
\end{align}
where we used  Eqs.~\eqref{eq:he_expansion_01} and~\eqref{eq:he_expansion_02} from Appendix B in components:
    \begin{align}
        \Gamma h^{(0)}_\parallel &= \nu \left(\partial_1^x u_1^{(0)}+\partial_1^y u_1^{(1)}\right)\label{eq:he_parallel_psi_0}\\
        \Gamma h^{(0)}_\perp &= \dfrac{(\nu-1)}{2}\left( \partial_1^x u_2^{(0)}+ \partial_1^y u_2^{(1)}\right) + \dfrac{(\nu + 1)}{2}\left( \partial_2^x u_1^{(0)}+ \partial_2^y u_1^{(1)}\right)\label{eq:he_perp_psi_0}
    \end{align}

Finally, because the pressures $p^{(-1)}$ and $p^{(0)}$ are Lagrange multipliers, we redefine them as follows in order to simplify the stress components in Eqs.~\eqref{eq:total_stress_index_he_11_order_0_v2}-\eqref{eq:total_stress_index_he_21_order_0_v2}:
\begin{align}
    \bar{p}^{(-1)}&=p^{(-1)}-\dfrac{\nu(\nu+1)}{2\Gamma}\partial_1 ^y u_1^{(0)}= p^{(-1)}+\dfrac{\nu(\nu+1)}{2\Gamma}\partial_2 ^y u_2^{(0)}\\
    \bar{p}^{(0)}&=p^{(0)}-\dfrac{\nu(\nu+1)}{2\Gamma}(\partial_1 ^x u_1^{(0)}+\partial_1 ^y u_1^{(1)})=p^{(0)}+\dfrac{\nu(\nu+1)}{2\Gamma}(\partial_2 ^x u_2^{(0)}+\partial_2 ^y u_2^{(1)})
\end{align}
where the equality follows from the incompressibility conditions Eqs.~\eqref{eq:forcebalance_expansion_23} and~\eqref{eq:forcebalance_expansion_13}.  Then, the stress components in Eqs.~\eqref{eq:total_stress_index_he_11_order_-1_v2}-\eqref{eq:total_stress_index_he_21_order_-1_v2} 
\begin{align}
    \sigma_{11}^{(-1)} &= \left(2\eta+\dfrac{\nu^2-\nu}{2\Gamma}\right) \partial_1^y u_1^{(0)} - \bar p^{(-1)} \label{eq:total_stress_index_he_11_order_-1_v3}\\
    \sigma_{22}^{(-1)} &= \left(2\eta+\dfrac{\nu^2+\nu}{2\Gamma}\right) \partial_2^y u_2^{(0)} - \bar p^{(-1)}\label{eq:total_stress_index_he_22_order_-1_v3}\\
    \sigma_{12}^{(-1)} &= (\eta+\dfrac{\nu^2-1}{4\Gamma}) \partial_1^y u_2^{(0)}+(\eta+\dfrac{(\nu+1)^2}{4\Gamma}) \partial_2^y u_1^{(0)}\label{eq:total_stress_index_he_12_order_-1_v3}\\
    \sigma_{21}^{(-1)} &= (\eta+\dfrac{(\nu-1)^2}{4\Gamma}) \partial_1^y u_2^{(0)}+(\eta+\dfrac{\nu^2-1}{4\Gamma}) \partial_2^y u_1^{(0)}\label{eq:total_stress_index_he_21_order_-1_v3}
\end{align}
and Eqs.~\eqref{eq:total_stress_index_he_11_order_0_v2}-\eqref{eq:total_stress_index_he_21_order_0_v2} become
\begin{align}
    \sigma_{11}^{(0)} &= \left(2\eta+\dfrac{\nu^2-\nu}{2\Gamma}\right) \left(\partial_1^x u_1^{(0)}+\partial_1^y u_1^{(1)}\right) - \bar p^{(0)} \label{eq:total_stress_index_he_11_order_0_v3}\\
    \sigma_{22}^{(0)} &= \left(2\eta+\dfrac{\nu^2+\nu}{2\Gamma}\right)\left(\partial_2^x u_2^{(0)}+\partial_2^y u_2^{(1)}\right) - \bar p^{(0)}\label{eq:total_stress_index_he_22_order_0_v3}\\
    \sigma_{12}^{(0)} &= (\eta+\dfrac{\nu^2-1}{4\Gamma}) \left(\partial_1^x u_2^{(0)} + \partial_1^x u_2^{(1)}\right)+(\eta+\dfrac{(\nu+1)^2}{4\Gamma}) \left(\partial_2^x u_1^{(0)} + \partial_2^x u_1^{(0)}\right)\label{eq:total_stress_index_he_12_order_0_v3}\\
    \sigma_{21}^{(0)} &= (\eta+\dfrac{(\nu-1)^2}{4\Gamma}) \left(\partial_1^x u_2^{(0)} + \partial_1^x u_2^{(1)}\right)+(\eta+\dfrac{\nu^2-1}{4\Gamma}) \left(\partial_2^x u_1^{(0)} + \partial_2^x u_1^{(0)}\right)\label{eq:total_stress_index_he_21_order_0_v3}
\end{align}

\section{Appendix D: Computing the force balance PDE}

In this Appendix, we use the results in Appendix C to perform the two-scale expansion for Eqs.~\eqref{eq: ericksen_leslie_psi_1}-\eqref{eq: ericksen_leslie_psi_4} as well as for the boundary conditions~\eqref{eq:BC_1}-\eqref{eq:BC_2} up to the first order corrections in $\eps$. 

Inserting the relations  \eqref{eq:two_scale_expansion} into Eqs.~\eqref{eq: ericksen_leslie_psi_1}-\eqref{eq: ericksen_leslie_psi_4} and Eqs.~\eqref{eq:BC_1}-\eqref{eq:BC_2} and treating the derivative as in Eq.~\eqref{eq:example}, one obtains to the lowest order in $\eps$ that
\begin{empheq}[left=\empheqlbrace]{align}
 \eps^{-2}:~~ &\eta_1 \partial_{11}^y u_1^{(0)}+\eta_2 
 \partial_{22}^y u_1^{(0)} - \partial_{1}^y 
 \bar p^{(-1)} = 0 &  x \in  \Omega^\eps\label{eq:forcebalance_expansion_21}\\
   \eps^{-2}:~~ &  \eta_1  \partial_{11}^y u_2^{(0)}+\eta_2 \partial_{22}^y u_2^{(0)} - \partial_{2}^y 
\bar p^{(-1)} = 0&  x \in  \Omega^\eps\label{eq:forcebalance_expansion_22}\\
   \eps^{-1}:~~ & \partial_{1}^y u_1^{(0)}+\partial_{2}^y u_2^{(0)}=0&  x \in  \Omega^\eps\label{eq:forcebalance_expansion_23}\\
   \eps^{-1}:~~ & \bs u ^{(0)} = \bs E \cdot \bs x&  x \in \partial \mathcal{P}^\eps \label{eq:forcebalance_expansion_24}\\
    \eps^{0}:~~& \bs u^{(0)}=\bs E\cdot \bs x &  x \in \partial \Omega
    \label{eq:forcebalance_expansion_25}
\end{empheq} 
where we used the stress components \eqref{eq:total_stress_index_he_11_order_-1}-\eqref{eq:total_stress_index_he_21_order_-1} in Appendix C. The parameters $\eta_1=\eta+(\nu-1)^2/4\Gamma$ and $\eta_2=\eta+(\nu+1)^2/4\Gamma$ are two effective viscosities. The solution of this set of PDEs is 
\begin{equation}
    \bs u^{(0)} = \bs E\cdot \bs x\label{eq:U0ref}
\end{equation}
Note that because the strain rate $\bs E$ represents a simple shear flow, it is traceless and symmetric. Therefore $\partial_1^x u_1^{(0)}+\partial_2^x u_2^{(0)}=0$.

Since the velocity and pressure fields are $\eps$-periodic in the fast variable $\bs y$, the corrections to the next order in $\eps$ of Eqs.~\eqref{eq: ericksen_leslie_psi_1}-\eqref{eq: ericksen_leslie_psi_4} and Eqs.~\eqref{eq:BC_1}-\eqref{eq:BC_2} are
\begin{empheq}[left=\empheqlbrace]{align}
 \eps^{-1}:~~ &\eta_1 \partial_{11}^y u_1^{(1)}+\eta_2 \partial_{22}^y u_1^{(1)} - \partial_{1}^y 
 \bar p^{(-1)} = 0 &  y \in  Y_f\label{eq:forcebalance_expansion_11}\\
   \eps^{-1}:~~ &  \eta_1 \partial_{11}^y u_2^{(1)}+\eta_2\partial_{22}^y u_2^{(1)} - \partial_{2}^y 
 \bar p^{(-1)} = 0&  y \in  Y_f\label{eq:forcebalance_expansion_12}\\
   \eps^{0}:~~ & \partial_{1}^y u_1^{(1)}+\partial_{2}^y u_2^{(1)}=0&  y \in  Y_f\label{eq:forcebalance_expansion_13}\\
   \eps^{0}:~~ & \bs u^{(1)}= -\bs E \cdot \bs y &  y \in \partial \mathcal{P}\label{eq:forcebalance_expansion_14}\\
    \eps^{1}:~~& \bs u^{(1)} \text{ periodic} &  y \in \partial \partial Y_f
    \label{eq:forcebalance_expansion_15}
\end{empheq} 

Because the fluid is incompressible, the velocity field can be expressed in terms of a streamfunction such that $u^{(1)}_1=-\partial^y_2\Psi^{(1)}$ and $u^{(1)}_2=\partial^y_1\Psi^{(1)}$. Taking the curl of~\eqref{eq:forcebalance_expansion_11} and~\eqref{eq:forcebalance_expansion_12}, one obtains that $\Psi^{(1)}$ solves an anisotropic biharmonic equation
 \begin{equation}
(\eta_1 \partial_{11}^y+\eta_2 \partial_{22}^y)(\partial_{11}^y+\partial_{22}^y)\Psi^{(1)}= 0\label{eq:anisotrpoic_biharmonic}
 \end{equation}
 Taking the divergence of~\eqref{eq:forcebalance_expansion_11} and~\eqref{eq:forcebalance_expansion_12}, one obtains that $\bar{p}^{(0)}$ solves the Laplace equation 
  \begin{equation}
(\partial_{11}^y+\partial_{22}^y)\bar{p}^{(0)}= 0\label{eq:Laplace_Pressure}
 \end{equation}
 The set of PDEs \eqref{eq:forcebalance_expansion_11}-\eqref{eq:forcebalance_expansion_15} is solved using asymptotic expansions in Appendix E.

\section{Appendix E: Computing the Weak Anisotropic Limit:}
In this Appendix, we perform the asymptotic expansion of the physical fields in the limits $\nu \ll 1 $ and $\nu \gg 1$ in the anisotropic Stokes Eqs.~\eqref{eq:forcebalance_expansion_11}-\eqref{eq:forcebalance_expansion_15} obtained in Appendix D.  This reduces the PDEs to a set of isotropic Stokes equations with forcing terms. 

We start from expanding the velocity and pressure field in the anisotropic Stokes Eq.~\eqref{eq:PDEu1_1}-\eqref{eq:PDEu1_3} as a power series in small shear-flow alignment $|\nu| \ll 1$~\eqref{eq:smallnu_power_series_expansion}
\begin{empheq}[left=\empheqlbrace]{align}
    \bs u^{(1)}_s &= \sum_{k=0}^\infty \nu^k \bs u^{(1,k)}_s\\
    \bar p^{(0)}_s &= \sum_{k=0}^\infty \nu^k \bar p^{(0,k)}_s
\end{empheq}
where we denote the $\eps$-order $\alpha$ and $\nu$-order $\beta$ function by $\bs u^{(\alpha,\beta)}$.  We collect terms at each order $\nu^{k}$
\begin{align}
    \nu^0 &:  \left\{
    \begin{aligned}
        &\bar \eta \Delta \bs u^{(1,0)}_s - \nabla \bar p ^{(0,0)}_s = 0\\
        &\nabla \cdot \bs u^{(1,0)}_s = 0
    \end{aligned}
    \right. & y \in Y_f\label{eq:nu_expansion_order_0}\\
    \nu^1 &: \left\{ 
    \begin{aligned}
        &\bar \eta \Delta \bs u^{(1,1)}_s - \nabla \bar p^{(0,1)}_s + \dfrac{1}{2\Gamma} \left(\partial_{22}^y \bs u^{(1,0)}_s - \partial_{11}^y \bs u^{(1,0)}_s
        \right)= 0\\
        &\nabla \cdot \bs u^{(1,1)}_s = 0
    \end{aligned}
    \right.& y \in Y_f\label{eq:nu_expansion_order_1}\\
    \nu^2 &: \left\{ 
    \begin{aligned}
        &\bar \eta \Delta \bs u^{(1,2)}_s - \nabla \bar p^{(0,2)}_s + \dfrac{1}{2\Gamma} \left(\partial_{22}^y \bs u^{(1,1)}_s - \partial_{11}^y \bs u^{(1,1)}_s
        \right) \\
        &\hspace{45pt} + \dfrac{1}{4\Gamma}\left(\partial_{11}^y \bs u^{(1,0)}_s + \partial_{22}^y \bs u^{(1,0)}_s\right)= 0\\
        &\nabla \cdot \bs u^{(1,2)}_s = 0
\end{aligned}\right.& y \in Y_f\label{eq:nu_expansion_order_2}
\end{align}
where $\bar \eta = \eta + 1/4\Gamma$.  We solve this system by a streamfunction $\Psi^{(1,k)}_s$
\begin{equation}
    \Psi^{(1,k)}_s := A^{(k)}(r) \sin(2(k+1)\theta) \label{eq:streamfunction_ansatz_small}
\end{equation}

Furthermore, we expand Eq.~\eqref{eq:PDEu1_1}-\eqref{eq:PDEu1_3} as a power series in large shear-flow alignment $|\nu| \gg 1$.
From Eq.~\eqref{eq:largenu_power_series_expansion}
\begin{empheq}[left=\empheqlbrace]{align}
    \bs u^{(1)}_l &= \sum_{k=1}^\infty \nu^{-k} \bs u^{(1,k)}_l\\
    \bar p^{(0)}_l &= \sum_{k=0}^\infty \nu^{-k} \bar p^{(0,k)}_l
\end{empheq}

\begin{align}
    \nu^2 &: \left\{ 
    \begin{aligned}
        &\bar \eta \Delta \bs u^{(1,2)}_l - \nabla \bar p^{(0,2)}_l = 0 \\
        &\nabla \cdot \bs u^{(1,2)}_l = 0
\end{aligned}\right.& y \in Y_f\label{eq:largenu_expansion_order_0}\\
\nu^1 &: \left\{ 
    \begin{aligned}
        &\bar \eta \Delta \bs u^{(1,1)}_l - \nabla \bar p^{(0,1)}_l + \dfrac{1}{2\Gamma} \left(\partial_{22}^y \bs u^{(1,1)}_l - \partial_{11}^y \bs u^{(1,1)}_l
        \right)= 0 \\
        &\nabla \cdot \bs u^{(1,1)}_l = 0
\end{aligned}\right.& y\in Y_f\label{eq:largenu_expansion_order_1}\\
\nu^0 &: \left\{ 
    \begin{aligned}
        &\bar \eta \Delta \bs u^{(1,0)}_l - \nabla \bar p^{(0,0)}_l + \dfrac{1}{2\Gamma} \left(\partial_{22}^y \bs u^{(1,1)}_l - \partial_{11}^y \bs u^{(1,1)}_l
        \right)= 0 \\
        &\hspace{45pt} + \dfrac{1}{4\Gamma}\left(\partial_{11}^y \bs u^{(1,2)}_l + \partial_{22}^y \bs u^{(1,2)}_l\right)= 0\\
        &\nabla \cdot \bs u^{(1,0)}_l = 0
\end{aligned}\right.& y \in Y_f\label{eq:largenu_expansion_order_2}
\end{align}
where we use the streamfunction $\Psi^{(1,k)}_l$
\begin{equation}
     \Psi^{(1,k)}_l := B^{(k)}(r) \sin(2(k+1)\theta) \label{eq:streamfunction_ansatz_large}, \quad k=0,1,2
\end{equation}
Using the scaling~\eqref{eq:diluteness_parameter}, we pass to the dilute limit in both~\eqref{eq:nu_expansion_order_0}-\eqref{eq:nu_expansion_order_2}  and~\eqref{eq:largenu_expansion_order_0}-\eqref{eq:largenu_expansion_order_2} by taking the particle radius $a \to 0$ as $\eps \to 0$, see Fig.~\ref{fig:fig2}.  In this limit, we rescale the domain $Y_f$ such that it becomes an infinite domain $\R^2\setminus \mathcal{P}$ with a particle of size $1$ and we replace the periodic boundary condition with a decay condition $\bs u^{(1)}\to 0$ at infinity.  Solving for the coefficients $A^{(k)}, B^{(k)}$ in the streamfunction ansatze~\eqref{eq:streamfunction_ansatz_small} and \eqref{eq:streamfunction_ansatz_large} in the infinite domain $\R^2\setminus \mathcal{P}$, we obtain~\eqref{eq:streamfunctionsmallnu} and \eqref{eq:streamfunctionlargenu}, respectively, after rescaling back to the unit cell domain $Y_f$. 

We solve explicitly the velocity field $\bs u^{(1)}$ by finding for either streamfunction $\Psi^{(1,k)}_s$ or $\Psi^{(1,k)}_l$
\begin{equation}
    u^{(1,k)}_1=-\partial^y_2\Psi^{(1,k)}, \quad u^{(1,k)}_2=\partial^y_1\Psi^{(1,k)}, \quad k=0,1,2
\end{equation}
Due to the incompressibility constraint the pressure satisfies a  Laplace equation at each order 
\begin{equation}
    \Delta \bar p^{(0,k)}=0, \quad k=0,1,2
\end{equation}

Note that for the 2D isotropic Stokes Eqs.~\eqref{eq:nu_expansion_order_0}-\eqref{eq:nu_expansion_order_2}  and Eqs.~\eqref{eq:largenu_expansion_order_0}-\eqref{eq:largenu_expansion_order_2}, the solution to the problem with rigidity condition~\eqref{eq:BC_1} coincides (see for example \cite{Gua2011}, Section 2) with the solution with a no-slip condition boundary condition 
\begin{equation}
    \bs u = 0
\end{equation}
where the velocity of the particle is assumed to be $0$, see Appendix F.

\section{Appendix F: Net force and net torque on the particle boundary:}

In this Appendix, we determine the net force and net torque on the particle boundary as a result of the perturbations of the physical fields. For these calculations, we will use the result from Appendix B to E. For more details on the derivation of the net force and net torque on the interface of a liquid crystal, we refer to \cite{furthauer2012taylor,julicher2018hydrodynamic}.

Because the particles are undeformable, their motion is limited to translations and rotations. The linear and angular velocities of a particle are determined by enforcing that the net force and the net torque on the particle boundary vanishes. 

On the interface of a liquid crystal and in the absence of external sources, the expressions of the net force and net torque are
\begin{align}
    &\int_{\partial \mathcal{P}^\eps}  \bs \sigma^\eps \cdot \hat{\bs N} dl = 0  \label{eq:totalfrocebalance}\\
    &\int_{\partial \mathcal{P}^\eps}  \bs r \times (\bs \sigma^\eps_d \cdot \hat{\bs N}) dl = 0  \label{eq:totaltorquebalance}
\end{align}
where $\bs \sigma^\eps$ is the total stress tensor, $\hat{\bs N}$ is the outward facing normal vector, and
\begin{align}
    &\bs \sigma^\eps_d=  2\eta D(\bs u^\eps)  +\dfrac{\nu}{2} (\bs n^\eps \bs h^\eps + \bs h^\eps \bs n^\eps) + \half (\bs n^\eps \bs h^\eps - \bs h^\eps \bs n^\eps)\label{eq:dev_stress}
\end{align}
and $dl$ is an infinitesimal element of length on the boundary, which in our case is the boundary of a particle.

The two-scale expansion of the stress components of $\bs \sigma^\eps$ and \eqref{eq:dev_stress} can be determined using the results in Appendix C. To order $\eps^0$, the components of $\bs \sigma^\eps$ are given by Eqs.~\eqref{eq:total_stress_index_he_11_order_0_v3}-\eqref{eq:total_stress_index_he_21_order_0_v3} and the components of takes a similar form
\begin{align}
    \sigma_{d,11}^{(0)} &= (2\eta+\dfrac{\nu^2}{\Gamma}) \left(\partial_1^x u_1^{(0)}+\partial_1^y u_1^{(1)}\right) \label{eq:Dev_stress_index_he_11_order_0_v2}\\
    \sigma_{d,22}^{(0)} &= 2\eta \left(\partial_2^x u_2^{(0)}+\partial_2^y u_2^{(1)}\right) \label{eq:Dev_stress_index_he_22_order_0_v2}\\
    \sigma_{d,12}^{(0)} &= (\eta+\dfrac{\nu^2-1}{4\Gamma}) \left(\partial_1^x u_2^{(0)} + \partial_1^x u_2^{(1)}\right)+(\eta+\dfrac{(\nu+1)^2}{4\Gamma}) \left(\partial_2^x u_1^{(0)} + \partial_2^x u_1^{(0)}\right)\label{eq:Dev_stress_index_he_12_order_0_v2}\\
    \sigma_{d,21}^{(0)} &= (\eta+\dfrac{(\nu-1)^2}{4\Gamma}) \left(\partial_1^x u_2^{(0)} + \partial_1^x u_2^{(1)}\right)+(\eta+\dfrac{\nu^2-1}{4\Gamma}) \left(\partial_2^x u_1^{(0)} + \partial_2^x u_1^{(0)}\right)\label{eq:Dev_stress_index_he_21_order_0_v2}
\end{align}
Using the expression for $\bs u^{(0)}$ in Eq.~\eqref{eq:U0ref} and the expression for $\bs u^{(1)}$ in Eqs.~\eqref{eq:streamfunctionsmallnu} or ~\eqref{eq:streamfunctionlargenu}, one can show that the net force and net torque vanish in the two limiting cases controlled by $\nu$. As a consequence the particle velocity vanishes.

\section{Appendix G: Macroscopic field of an homogenized medium:}

In this Appendix, we determine the form of the physical fields of an homogenized liquid crystal medium that is subjected to the same extensional shear flows and preferred anchoring as the suspension \eqref{eq:BC_5} and \eqref{eq:BC_6}. 

When the homogenized medium is subjected to an external shear flow, each particle can create a disturbance on the macroscopic scales that can influence the and the integrate effect can generate corrections to the macroscopic flows. Following \cite{Ein1906}, we consider that the flow field of the homogenized medium is  
\begin{align}
\bs u^H&=\bs E\cdot\bs x+\sum_{n=1}^{N_p}\bs u^{(1)}(\bs x,\bs y=\bs x-\bs x_i)    
\end{align}
where the summation runs over all particles in the medium and $\bs x_i$ is the center of the i-th particle. Because in our case, particles do not perturb the director field (see Appendix B), the corrections of an ensemble of particles to the macroscopic director field vanishes. 

The corrected macroscopic shear flow is defined as
\begin{align}
E^H_{\alpha\beta}&= E_{\alpha\beta}+\left(\sum_{n=1}^{N_p}  \partial_{x_{\alpha}}(u^{(1)}_\beta(\bs x,\bs y=\bs x-\bs x_i))\right)|_{\bs x=0}    \label{eq:shearM1}
\end{align}
where the first term on the RHS represents the macroscopic flows and the second term on the RHS represents the sum of the disturbance flows generated by each particle. Considering that the particles are evenly distributed and taking the continuum limit, we can express \eqref{eq:shearM1} as
\begin{align}
E^H_{\alpha\beta}&=E_{\alpha\beta}+\frac{N_p}{|\Omega|}\int_\Omega \partial_{x_\alpha}(u^{(1)}_\beta(\bs x,\bs y=\bs x-\bs x'))|_{\bs x=0} d\bs x'\label{eq:shearM2}\\
&=E_{\alpha\beta}-\frac{N_p}{|\Omega|}\int_\Omega u^{(1)}_\beta(\bs x,\bs y=\bs x-\bs x')|_{\bs x=0} x'_{\alpha} d\bs x'\label{eq:shearM3}
\end{align}
Substituting the expression of the perturbation flows $u^{(1)}_\alpha$ in the two limits of $\nu$, Eqs.~\eqref{eq:streamfunctionsmallnu} or ~\eqref{eq:streamfunctionlargenu}, leads to the same expression for the homogenized macroscopic shear flow 
\begin{equation}
        E^H=\left( \begin{array}{cc}
       \gamma(1-\phi)  & 0\\
        0 & -\gamma(1-\phi)\end{array} \right)\label{eq:physicslfieldH}
\end{equation}

\end{document}